\documentclass[10pt,letterpaper]{article}
\usepackage[top=0.85in,left=2.75in,footskip=0.75in]{geometry}

\usepackage{changepage}

\usepackage[utf8x]{inputenc}

\usepackage{textcomp,marvosym}

\usepackage{amsmath,amssymb}

\usepackage{cite}

\usepackage{nameref,hyperref}

\usepackage[right]{lineno}

\usepackage{microtype}
\DisableLigatures[f]{encoding = *, family = * }

\raggedright
\usepackage[aboveskip=1pt,labelfont=bf,labelsep=period,justification=raggedright,singlelinecheck=off]{caption}

\makeatletter
\renewcommand{\@biblabel}[1]{\quad#1.}
\makeatother

\date{}

\usepackage{lastpage,fancyhdr,graphicx}
\usepackage{epstopdf}
\fancyheadoffset[L]{2.25in}
\fancyfootoffset[L]{2.25in}
\begin{document}
\vspace*{0.2in}

\begin{flushleft}
{\Large
\textbf\newline{Statistical Models for Tornado Climatology: Long and Short-Term Views} 
}
\newline
\\
James B. Elsner\textsuperscript{1*},
Thomas H. Jagger\textsuperscript{1}, and
Tyler Fricker\textsuperscript{1}
\\
\bigskip
\textbf{1} Department of Geography, Florida State University, Tallahassee, Florida, USA
\\
\bigskip

%
%



* jelsner@fsu.edu

\end{flushleft}
\section*{Abstract} 
This paper estimates local tornado risk from records of past events using statistical models. First, a spatial model is fit to the tornado counts aggregated in counties with terms that control for changes in observational practices over time. Results provide a long-term view of risk that delineates the main tornado corridors in the United States where the expected annual rate exceeds two tornadoes per 10,000 square km. A few counties in the Texas Panhandle and central Kansas have annual rates that exceed four tornadoes per 10,000 square km. Refitting the model after removing the least damaging tornadoes from the data (EF0) produces a similar map but with the greatest tornado risk shifted south and eastward. Second, a space-time model is fit to the counts aggregated in raster cells with terms that control for changes in climate factors. Results provide a short-term view of risk. The short-term view identifies the shift of tornado activity away from the Ohio Valley under El Ni\~no conditions and away from the Southeast under positive North Atlantic oscillation conditions. The combined predictor effects on the local rates is quantified by fitting the model after leaving out the year to be predicted from the data. The models provide state-of-the-art views of tornado risk that can be used by government agencies, the insurance industry, and the general public.\\

\noindent {\it Keywords: tornadoes, spatial model, risk prediction, climate, space-time model}


\section*{Introduction}

Seasonal climate forecasts are issued routinely. For example, each spring weather agencies in nations across the world make predictions for how hot and dry the summer is likely to be. And predictions for hurricane activity along the coast are typically accurate enough to warrant attention by the property insurance industry. However, there is yet no credible forecasts of tornado activity months in advance. This despite demonstrated useful skill (accuracy above random guess) at predicting tornado activity prior to the start of the season \cite{ElsnerWiden2014, AllenEtAl2015}. The absence of seasonal tornado forecasts is due to large gaps in the understanding of how climate affects severe weather and to the dearth of methods to forecast activity on this time scale. 

Tornadoes are too small to be resolved in the dynamical models that are routinely used as guidance in making seasonal forecasts. Predictions of the large-scale environments conducive to tornadoes can be made with dynamical models but the necessary conditions are not sufficient to distinguish between days with and without tornadoes. An alternative is to fit statistical models to historical data. Climate patterns related to active and inactive tornado seasons provide information to make such predictions but population growth and changes to procedures for rating tornadoes result in a heterogeneous database. Various methods for dealing with data artifacts have been proposed \cite{King1997, RayEtAl2003, AndersonEtAl2007} with most procedures assuming a uniform region of activity and estimating occurrence rates within a subset of the region likely to be most accurate. For example, tornado reports are often aggregated using kernel smoothing \cite{BrooksEtAl2003, DixonEtAl2011, ShaferDoswell2011}. The resulting spatial density maps show regions of higher and lower tornado frequency that are useful for exploratory analysis and hypothesis generation. Correctly interpreting the patterns is a problem however since there is no control for environmental factors. Another drawback is the assumption (implicit) that tornadoes occur randomly (not clustered). This is not generally the case as a single thunderstorm can spawn a family of tornadoes within a relatively compact area \cite{DoswellBurgess1988}. Also, tornado reports tend to be more numerous near cities compared to rural areas confounding attempts at assessing the risk over large regions. Moreover, this spatial variation in report frequency is decreasing with time \cite{ElsnerMichaelsScheitlinElsner2013}. Improvements in observing practices tend to result in more tornado reports, especially reports of weak tornadoes \cite{DoswellEtAl2005, VerboutEtAl2006} and ones occurring over remote areas. Finally, natural climate variations make some seasons more active than others. For instance, variations in sea-surface temperature and atmospheric convection in the tropical Pacific associated with the El Ni\~no/Southern Oscillation (ENSO) modulate global weather and climate patterns including the threat of tornadoes \cite{MarzbanSchaefer2001, CookSchaefer2008, MunozEnfield2011, LeeEtAl2013, AllenEtAl2015}. In short, statistical models capable of controlling for these various factors and artifacts are the tools for assessing tornado risk.

The goal of this paper is to demonstrate the science and technology under-girding tornado risk assessments. The work follows closely that of \cite{ElsnerFrickerJaggerMesev2016} but expands this earlier work both in the larger spatial domain and in the number of predictors. Part one fits a climatology model to data aggregated by county in states across the Midwest, Plains, and Southeast (long-term view of risk). The model uses annual population to control for changes in observational practices over time. Results give a long-term view of risk independent of climate variability. Part two fits a conditional climatology model to data aggregated in grid cells that predicts how the rates should be adjusted given current (or projected) climate conditions. The model controls for changes in observational practices over time using a trend term. Results give a short-term view of risk that depends on current climate conditions. A discussion of the model results and the potential utility of the models follows.

\section*{Materials and Methods}

\subsection*{Data}

A key variable in the long-term model is annual population by county that serves as a proxy for changes in observational practices. These values are archived by the U.S. Census Bureau and available from \url{www.nber.org/data/census-intercensal-county-population.html}. The latest cleaned population values are available for 2012. Since the population-effect has diminished over time and since population changes at the county level over a few years are small it is almost certain that using population only to 2012 will have no discernible influence on the results. We compute population density by dividing the population by county area and express the values in persons per square kilometers.

\begin{figure}[!h]
\caption{County population in 2012.}
\includegraphics[width=5in]{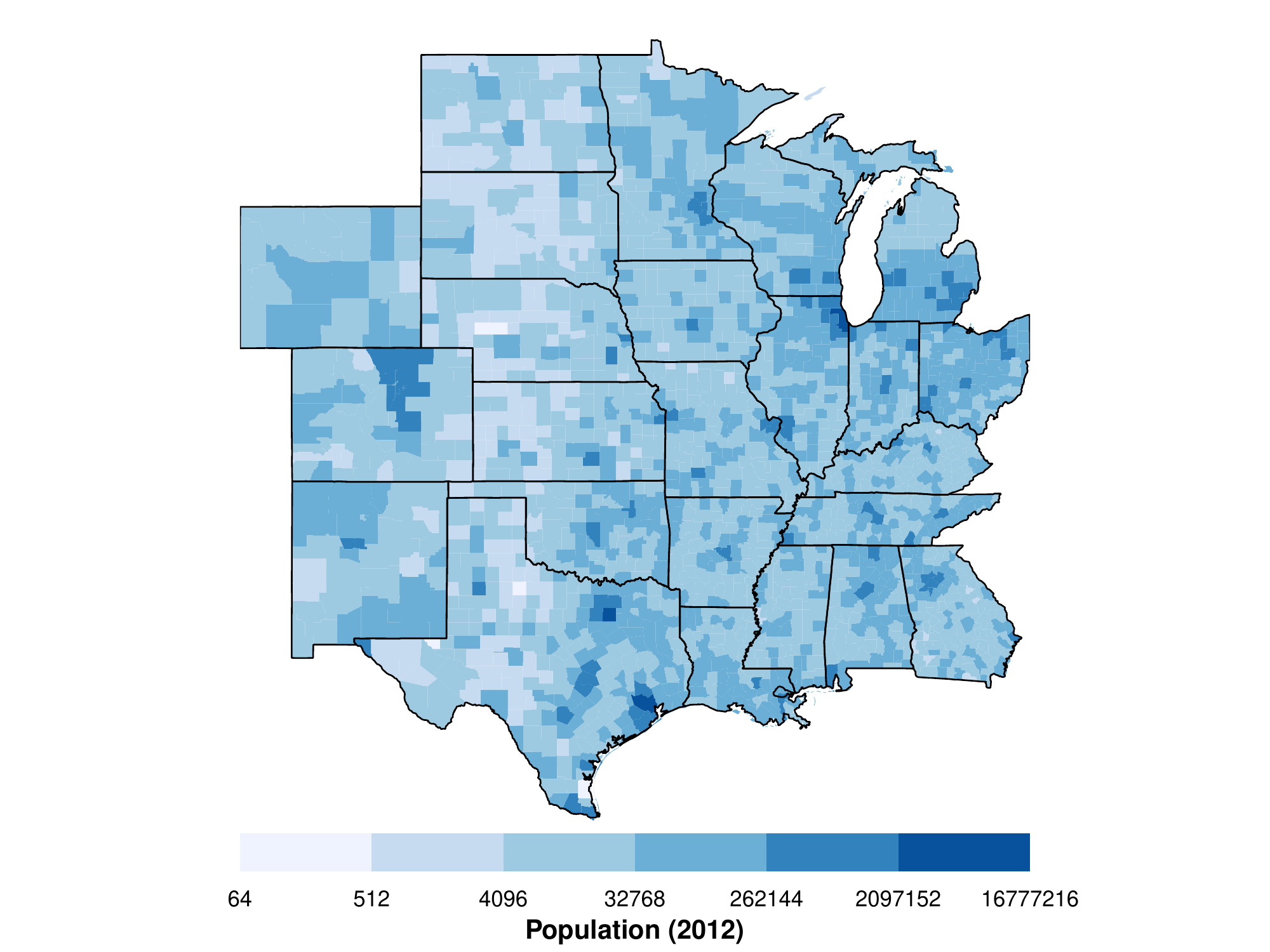}
\label{Population}
\end{figure}

The U.S. Census Bureau map boundaries are available from \url{www.census.gov/geo/maps-data/data/cbf/cbf_counties.html}. Here the 5m = 1:5,000,000 scale is used. State boundaries are extracted using the state Federal Information Processing Standard (FIPS). The set of 24 contiguous states including Wyoming, Colorado, New Mexico, North and South Dakota, Nebraska, Kansas, Oklahoma, Texas, Louisiana, Arkansas, Missouri, Iowa, Minnesota, Wisconsin, Michigan, Illinois, Indiana, Ohio, Kentucky, Tennessee, Georgian, Alabama, and Mississippi define the study area. There are 2168 counties across the two dozen states covering 7.6 million square kilometers. Combining the population data with the map boundaries the county-level population for 2012 is mapped in Fig.~\ref{Population}. A north-south region of low population density extends across the western portions of the high Plains.

The tornado data are from NOAA's Storm Prediction Center (SPC) and available from \url{www.spc.noaa.gov/gis/svrgis/zipped/tornado.zip}. The SPC maintains a comprehensive and up-to-date tornado database. Records extend back to 1950 and up through 2015 and include occurrence time, location, magnitude, track length and width, fatalities, and injuries for tornadoes in the United States. The version of the SPC database used in this study is in shapefile format with each tornado represented as a straight-line track in a Lambert conformal conic (LCC) projection centered on 96$^\circ$ W longitude and parallels at 33$^\circ$ and 45$^\circ$ N latitudes. The native coordinates are transformed to a Mercator projection.

\begin{figure}[!h]
\centering
\includegraphics[width=5in]{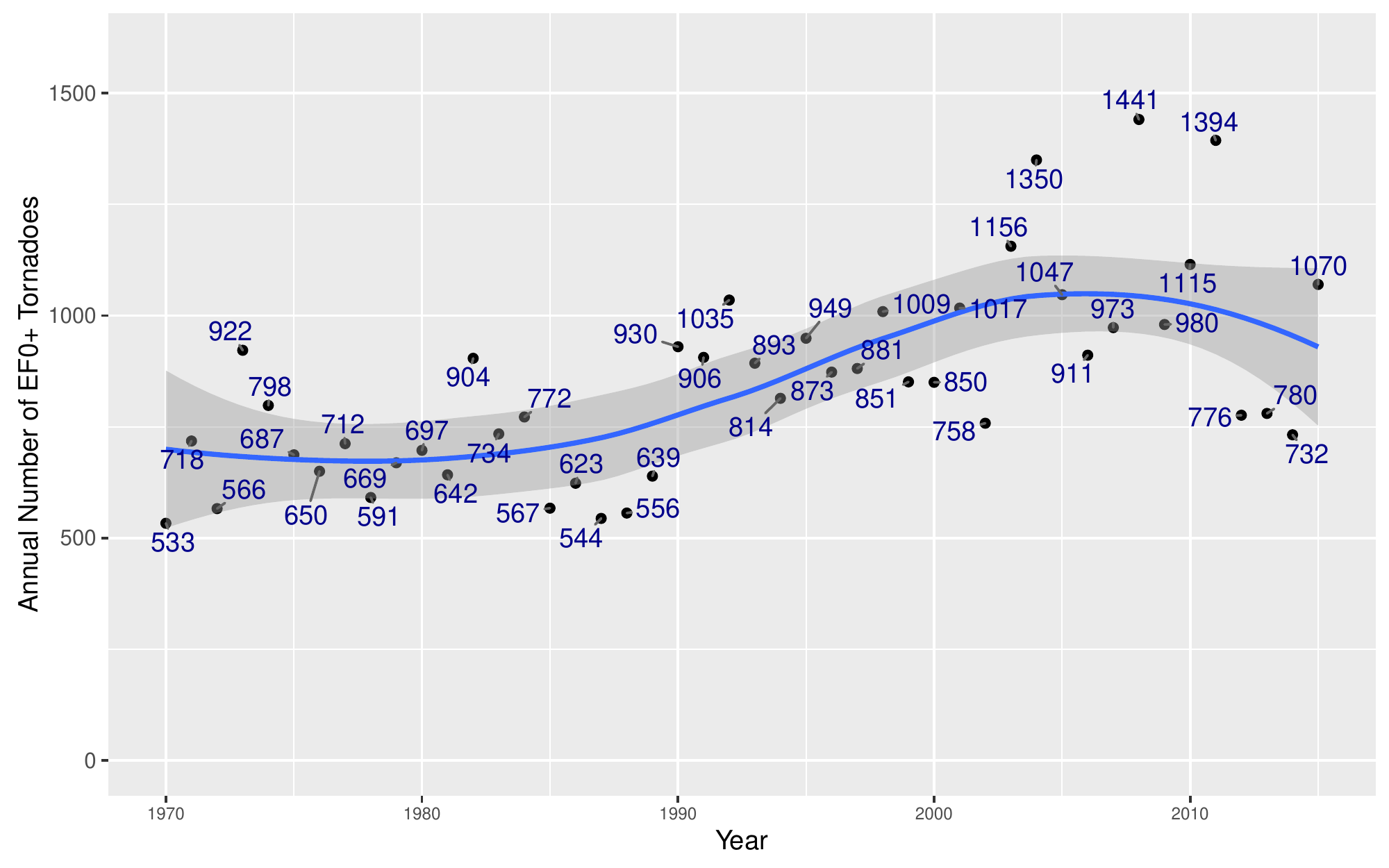}
 \caption{Annual number of tornadoes over the states used in the long-term view model.}
\label{TimeSeries}
\end{figure}

Tornado tracks are buffered to create damage-path polygons. The buffer is one half the value of the width variable specified in the attribute table. A flat cap on the buffer is used so the damage path is the same length as the track. The polygon paths are laid on top of the domain and a vector is returned indicating either NA (no portion of the damage path is inside the study area) or county numbers indicating which counties where affected.  Duplicate paths (1.03\% of all paths) are removed by checking for exact matches in width, length, date, time, and start location. For the long-term view all tornadoes starting with 1970 are used to create a 46-year climatology. There are a total of 39,015 tornadoes over this period and region. The annual number of tornadoes are plotted in Fig.~\ref{TimeSeries}. There is an upward trend in the annual counts that appears to have reversed around 2005. County level trends are included in the model.

\begin{figure}[!h]
\noindent \includegraphics[width=2.5in]{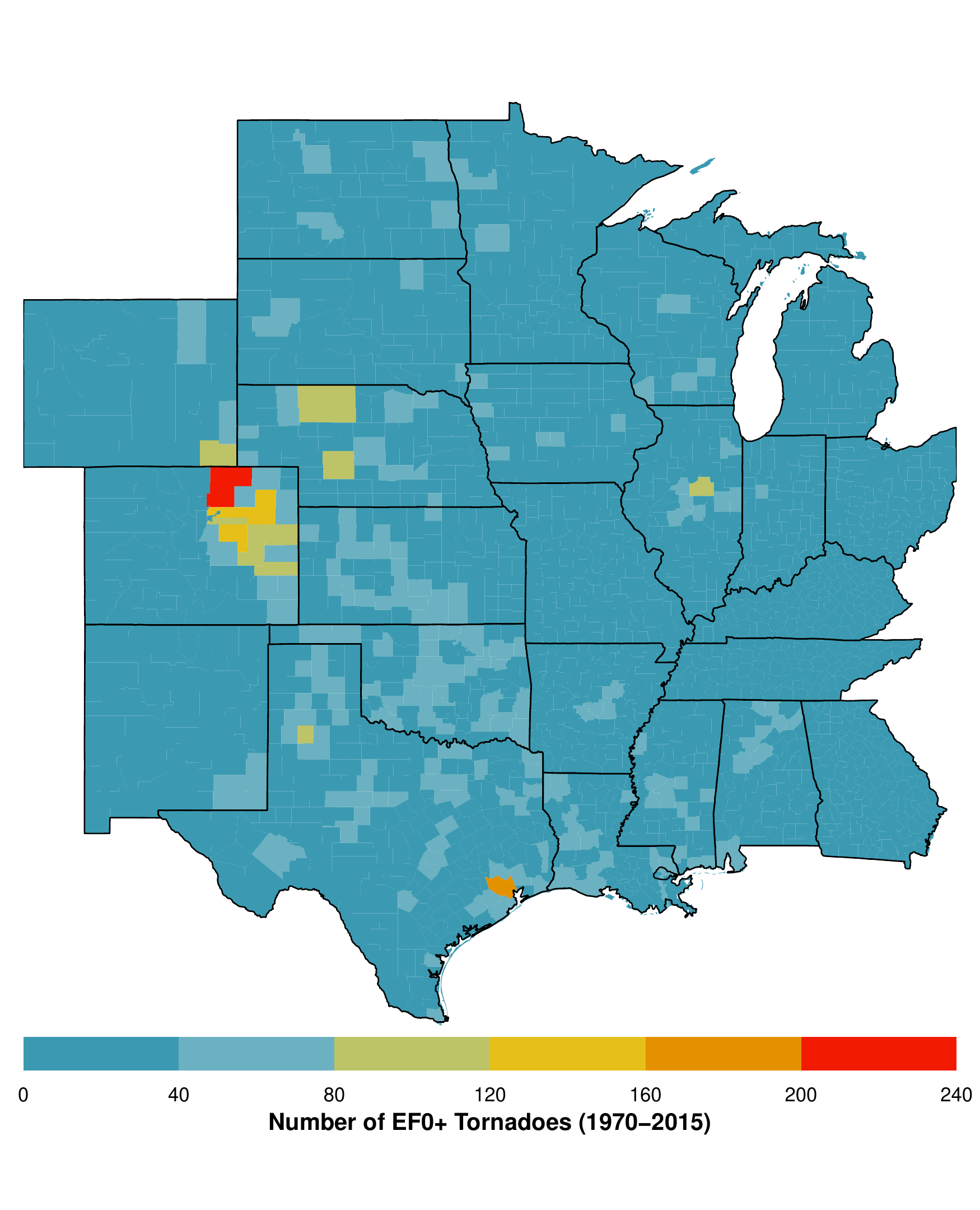}\includegraphics[width=2.5in]{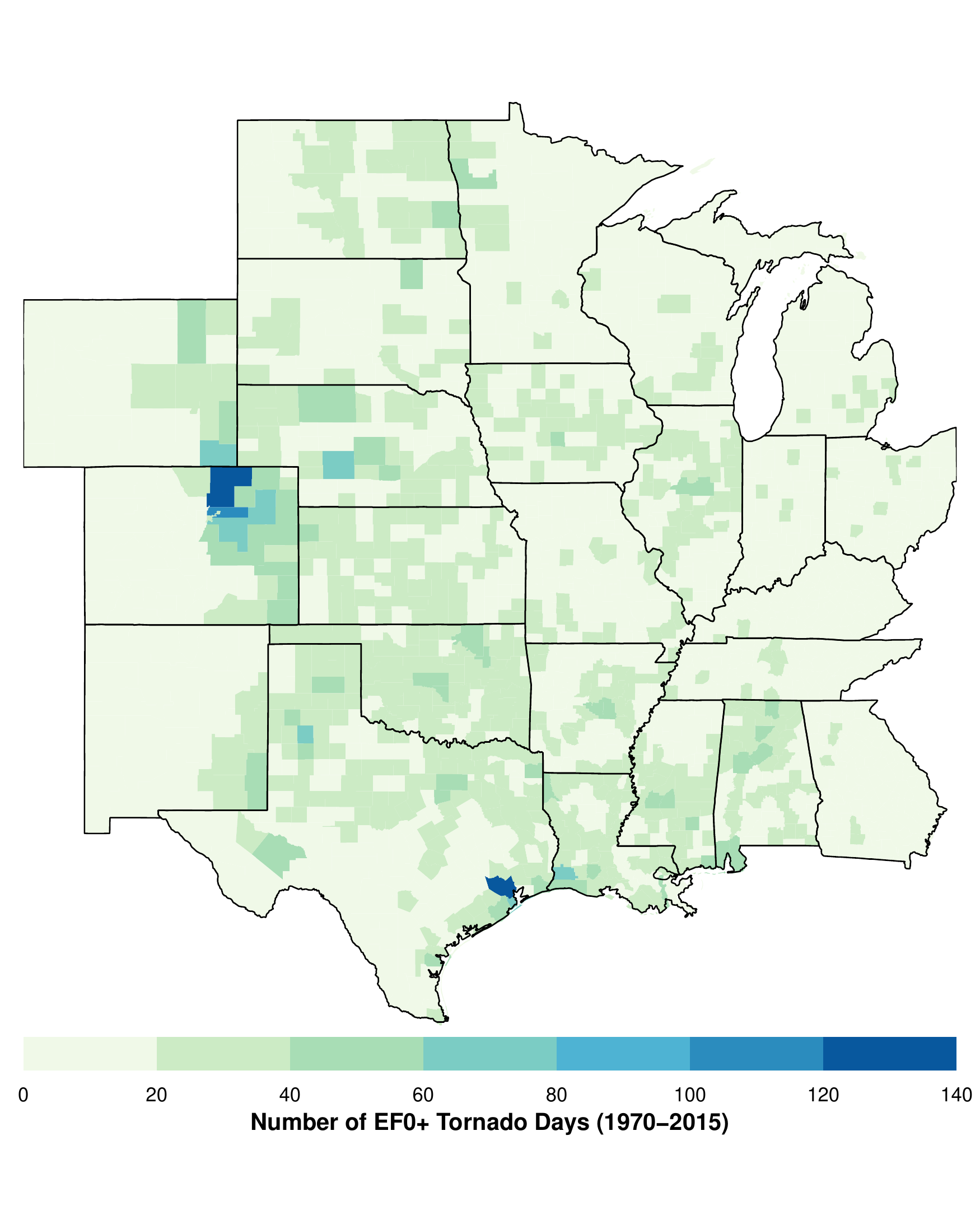}
 \caption{Number of tornadoes 1970--2015 (left) and number of days with at least one tornado.}
\label{nTnD}
\end{figure}

Paths are laid over the county boundaries to get a per-county tornado count. The result is a list of length equal to the number of counties with each element of the list containing a subset of the track attribute table. Padding with zeros is needed for the 19 counties without tornadoes (less than 1\%). Then for each county the number of tornadoes and tornado days are tabulated and plotted (Fig.~\ref{nTnD}). The maps show a tendency for larger counties to have more tornadoes and tornado days. For example, leading the list is Weld County, Colorado with a total of 219 tornado reports falling on 137 tornado days but spread over more than 18,000 square km. Next is Harris County, Texas with a total of 192 tornadoes falling on 124 days spread over more than 6,000 square km. County areas are included in the model (see next subsection) as a part of the exposure term.

\begin{figure}[!h]
\noindent \includegraphics[width=5in]{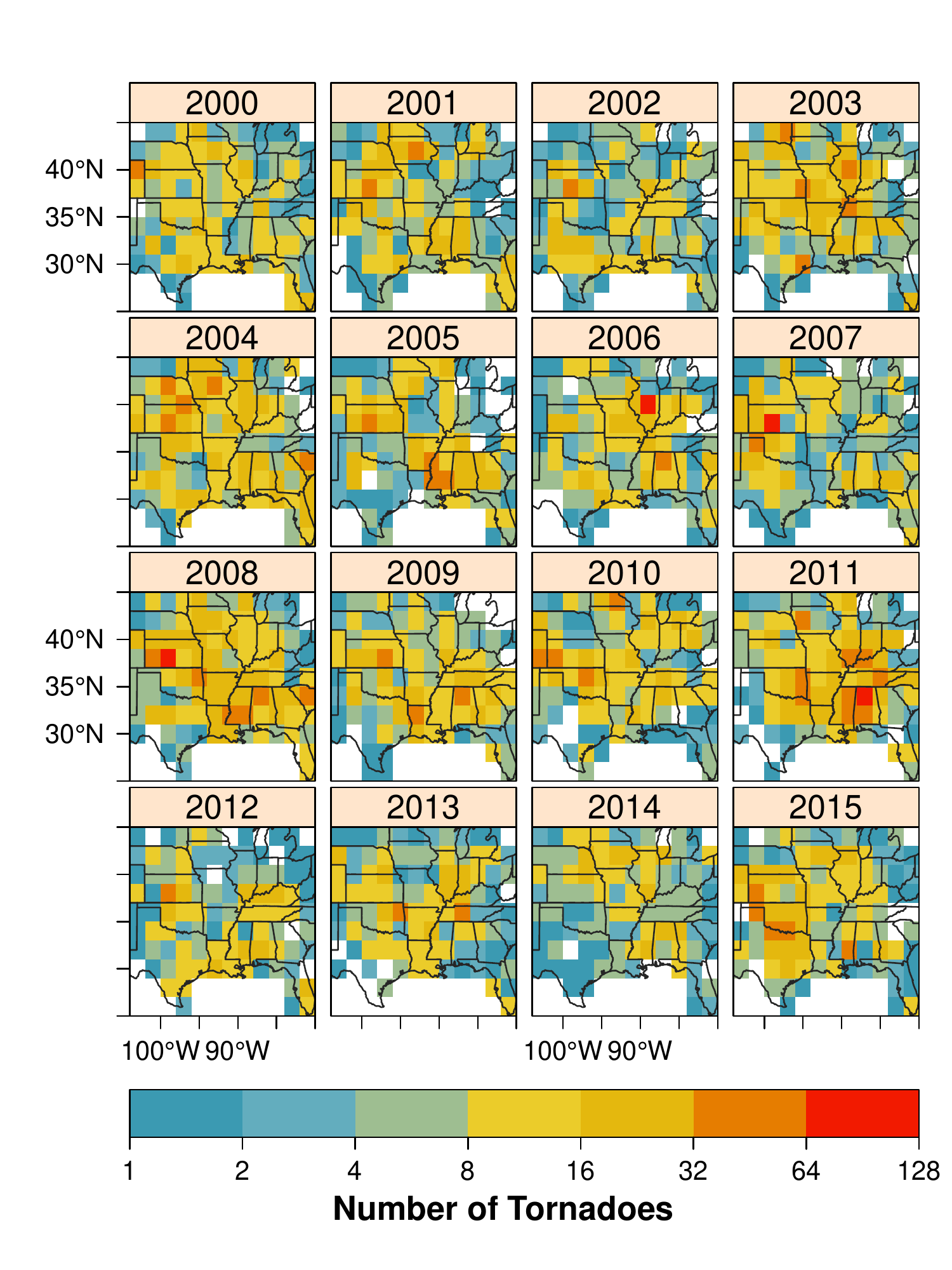}
 \caption{Annual tornado counts over the period 2000--2015 in two degree raster cells. A count is added to a cell if a tornado path intersects it.}
\label{Spacetime}
\end{figure}

A short-term view quantifies how much the tornado risk changes with a unit change in the climate variable (predictor). If the influence varies regionally then quantification is done spatially. To manage spatial variation, the short-term view model uses a raster of grid cells rather than a vector of counties. The raster has uniform size and shape contiguous cells making computations and comparisons easier. Annual tornado occurrences are counted in two degree cells based on track intersections (Fig.~\ref{Spacetime}). To simplify the overlay operation straight-line tracks are used instead of paths. The result is a space-time data set with cell area as a constant-time attribute and tornado count as a variable-time attribute. The study area extends from eastern Colorado to western Virginia and from the Mexican Gulf coast to southern Minnesota. The period of record runs from 1954 to 2015 for a total of 48,200 tornadoes representing 81.7\% of all tornadoes and 91.6\% of all violent tornadoes (EF4+). Data earlier than 1970 are included here since the model has a trend term for the increasing probability of tornado detection with increasing modernization. It is clear that at the cell level there is large variability in counts from one year to the next. Variability across space is also large in some years.

Key variables in the short-term view model are the climate predictors. Predictors are chosen based on literature research. They include indexes for ENSO and for the North Atlantic Oscillation (NAO) and sea-surface temperatures (SST) from the Gulf of Alaska and Western Caribbean Sea. The ENSO index, in units of standard deviation, is the bi-variate ENSO time series averaged from March through May. The monthly series combines a standardized Southern Oscillation Index with a standardized Ni\~no3.4 SST series computed from the Hadley Centre's data. The monthly values from March through May are averaged to obtain the index. Monthly NAO values, in units of standard deviation, are constructed from a rotated principal component analysis of the 500 hPa heights across the Northern Hemisphere. Higher than average heights over the subtropical Atlantic combined with lower than average heights in the vicinity of Iceland result in a strongly positive values of the NAO. Details of the procedure are given in \cite{BarnstonLivezey1987}. Monthly values from April through June are averaged to obtain the NAO index. 

\begin{figure}[!h]
\noindent \includegraphics[width=5in]{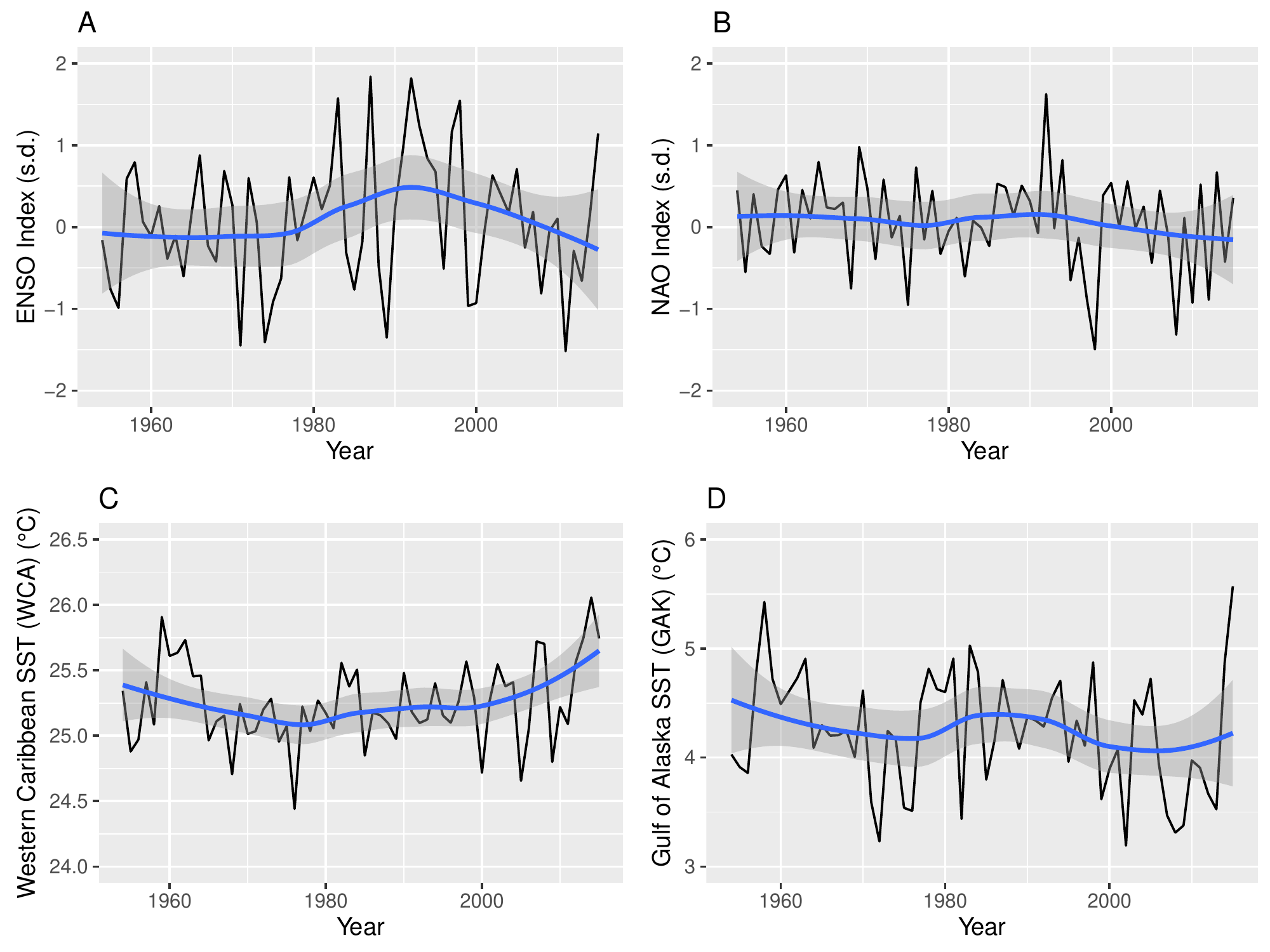}
 \caption{Time series of the four predictors used in the short-term view model.}
\label{Predictors}
\end{figure}

The SST regions are selected based on \cite{ElsnerWiden2014} who show that there is a tendency for a combination of colder than average ocean waters in the Gulf of Alaska and warmer than average ocean waters in the Western Caribbean Sea to favor tornado activity across the central United States during spring. The monthly SST values are spatially averaged from the NCEP/NCAR reanalysis grids \cite{KalnayEtAl1996}. The Gulf of Alaska region is bounded by 60 and 50.4$^\circ$ N latitudes and 136 and 153.6$^\circ$ W longitudes. Following \cite{ElsnerWiden2014} the GAK is the spatial average over the region for the month of April. The Western Caribbean Sea region is bounded by 25 and 15$^\circ$ N latitudes and 70 and 90$^\circ$ W longitudes. The WCA is the spatial average over the region for the month of February. The indexes and SST data are obtained from the Physical Science Division of the Earth System Research Laboratory. Time series plots (Fig.~\ref{Predictors}) of the four predictors used in the short-term view model indicate year-to-year variation but no significant upward or downward trends. 

\subsection*{Models}

Raw counts are problematic for directly assessing tornado rates because counties vary in size and population. To control for this a statistical model is fit to the counts. The long-term view the model includes population density as a fixed effect. To account for improvements in the procedures to rank tornadoes by the amount of damage, the calendar year and an interaction term of year with population are included. Mathematically, the number of tornadoes in each county $s$ ($T_{s})$ is assumed to be described by a negative binomial distribution with parameters probability $p$ and size $r$ \cite{ElsnerWiden2014}. If $X$ is a random sample from this distribution, then the probability that $X=k$ is $P(k|r,p) = \binom{k+r-1}{k} (1-p)^r p^k$, for $k \in 0,...,\infty$, $p \in (0, 1)$ and $r > 0$. This relates the probability of observing $k$ successes before the $r$-th failure of a series of independent events with probability of success equal to $p$. The distribution is generalized by allowing $r$ to be any positive real number and it arises from a Poisson distribution whose rate parameter can be described by a gamma distribution \cite{JaggerEtAl2015}.

The negative binomial distribution is re-formulated using the mean $\mu=r \frac{p}{1-p}$ and the size $r$. This separates the mean from the dispersion parameter. The mean $\mu_{s}$ is linked to a linear combination of the predictors and random terms, $\nu_{s}$ through the exponential function and the area of the cell, $A_s$ (exposure). The dispersion is modeled with a scaled size parameter $n$ where $n=r_{s}/A_s$ giving a dispersion of $1/p_{s}=1+\mu_{s}/n = 1 + \hbox{exp}(\nu_{s})/n$ that depends only on the tornado rate and $n$. More concisely the model is:
\begin{eqnarray}
T_{s} | \mu_{s}, r_{s}&\sim& \hbox{NegBin}(\mu_{s}, r_{s})   \\
\mu_{s} &=& A_s\exp(\nu_{s}) \\
\nu_{s} &=& \beta_0 + \beta_1\,\hbox{lpd}_{s} + \beta_2\,(t-t_0) + 
\beta_3\;\hbox{lpd}_{s}(t-t_0) + u_s + v_t \\
r_{s} &=& A_s \,n
\end{eqnarray}
where NegBin($\mu_{s}$, $r_{s}$) indicates that the conditional tornado counts ($T_{s}|\mu_{s}$, $r_{s}$) are described by negative binomial distributions with mean $\mu_{s}$ and size $r_{s}$,  $ \hbox{lpd}_{s}$ represents the base two logarithm of the population density during 2012 for each county, and $t_0$ is the base year set to 1991 (middle year of the record). The spatially correlated random term $u_s$ follows an intrinsic Besag formula with a sum-to-zero constraint \cite{Besag1975}.
\begin{eqnarray}
u_i | \{u_{j, j \neq i}, \tau \}&\sim& \mathcal{N}\left(\frac{1}{m_i} \sum_{i \sim j} u_j , \frac{1}{m_i} \tau \right),
\end{eqnarray}
where $\mathcal{N}$ is the normal distribution with mean $1/m_i \cdot \sum_{i \sim j} u_j$ and variance $1/m_i \cdot 1/\tau$ where $m_i$ is the number of neighbors of cell $i$ and $\tau$ is the precision; $i \sim j$ indicates cells $i$ and $j$ are neighbors. Neighboring cells are determined by contiguity (queen's rule). The annual uncorrelated random term, $v_t$, is modeled as a sequence of normally distributed random variables, with mean zero and variance $1/\tau'$. The prior on the spatial random term is statistically independent from the annual random term.

The short-term view model extends the long-term view model. Subscripts on parameters and variables indicate a space ($s$) {\it and} a time ($t$) component. Mathematically, the raw tornado count in grid cell $s$ for year $t$ is given as:
\begin{eqnarray}
T_{s,t} | \mu_{s, t}, r_{s,t}&\sim& \hbox{NegBin}(\mu_{s,t}, r_{s,t})   \\
\mu_{s,t} &=& A_s\exp(\nu_{s,t}) \\
\nu_{s,t} &=& \beta_1\,(t-t_0) + \beta_2\,\hbox{ID}_s + \beta_3\,\hbox{GAK}_t + \beta_{4,s}\;\hbox{ENSO}_t \nonumber \\
          &+& \beta_{5,s}\;\hbox{NAO}_t + \beta_{6,s}\;\hbox{WCA}_t\\ 
r_{s,t} &=& A_s \,n
\end{eqnarray}
where again the conditional tornado count in each cell is described by a negative binomial distribution with mean $\mu_{s,t}$ and size $r_{s,t}$. 

The annual rate in each cell $\mu_{s,t}$ is linked to a linear combination of the predictors ($\nu_{s,t}$) through the exponential function and the area of the cell, $A_s$. The predictors include the Gulf of Alaska SST (GAK) as a spatially-uniform effect and ENSO, NAO, and the Western Caribbean SST (WCA) as spatially-varying effects. As above, the spatially-varying effects have an intrinsic Besag formula (Eq.~5). Each cell is allowed to have unique variability through the ID term. The cell area times the number of years is the square-meter-years exposed to tornadoes and is normalized to have a mean of one over the domain.

Gaussian priors with low precision are assigned to the $\beta$'s. To complete the models the scaled size ($n$) is assigned a log-gamma prior and the precision parameters ($\tau$ and $\tau'$) are assigned a log-Gaussian prior \cite{JaggerEtAl2015}. Application of Bayes rule using the method of integrated nested Laplace approximation (INLA) \cite{RueMartinoChopin2009,RueEtAl2014} results in posterior densities for the model parameters. 

\section*{Results}

\subsection*{Long-Term View}
\subsubsection*{Climatology}

\begin{figure}[!h]
\noindent \includegraphics[width=5in]{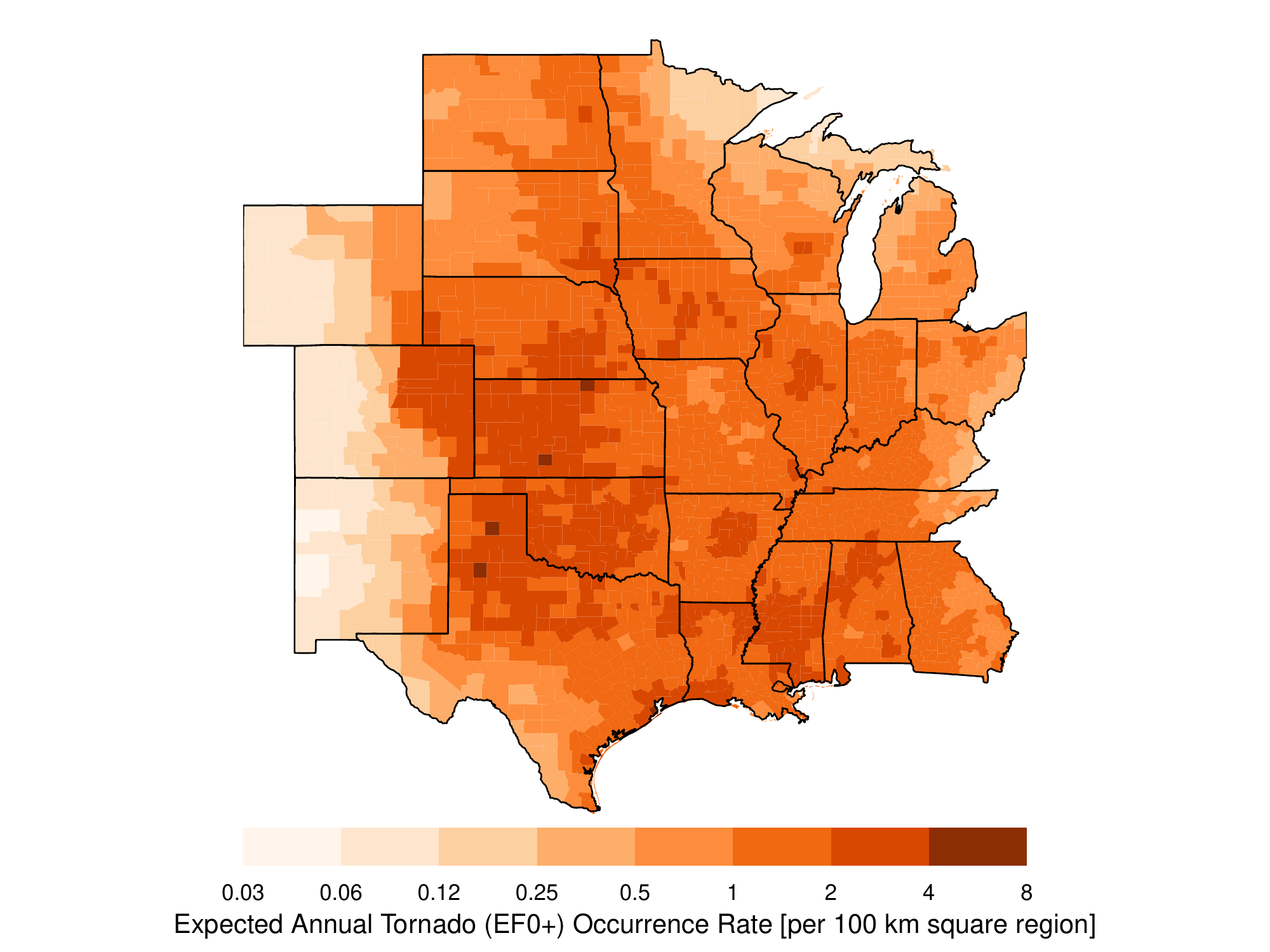}
 \caption{Long-term view. Expected annual tornado (EF0+) occurrence rates per 100 square kilometers.}
\label{OccurrenceRates}
\end{figure}

By specifying a `future' year (here 2016) the long-term model predicts a distribution for the counts in each county. The mean value of the predictive distribution is the expected annual occurrence rate conditional on the historical raw counts and the model. Values are standardized by dividing by county area and expressing them as a rate per 100 km square region (Fig.~\ref{OccurrenceRates}). This allows comparisons that are independent of county area. Note that the average county area is approximately 3500 km$^2$ so the per county rate in many cases is smaller. The color ramp is on a logarithmic scale so that each level of color saturation indicates a doubling of the occurrence rate.

As anticipated the broad-scale pattern of tornado activity matches the well-known climatology. Highest rates are found from the northern High Plains into the Mid South. Lowest rates are found over the Rockies, the Rio Grande Valley, and the northern Great Lakes. Regional features include lower rates over the Appalachian Mountains and a local minimum to the northeast of the Ozark Mountains. Occurrence rates are consistently above two tornadoes per year across the central and southern Great Plains. Rates drop off rapidly moving westward with values generally less than one tornado every eight years across the Rockies. Rates drop off less rapidly moving eastward with most regions exceeding one tornado per year with the exceptions of the northern parts of Minnesota, Wisconsin, and Ohio, and over the Appalachian regions of Ohio and Kentucky.

\begin{figure}[!h]
\noindent \includegraphics[width=5in]{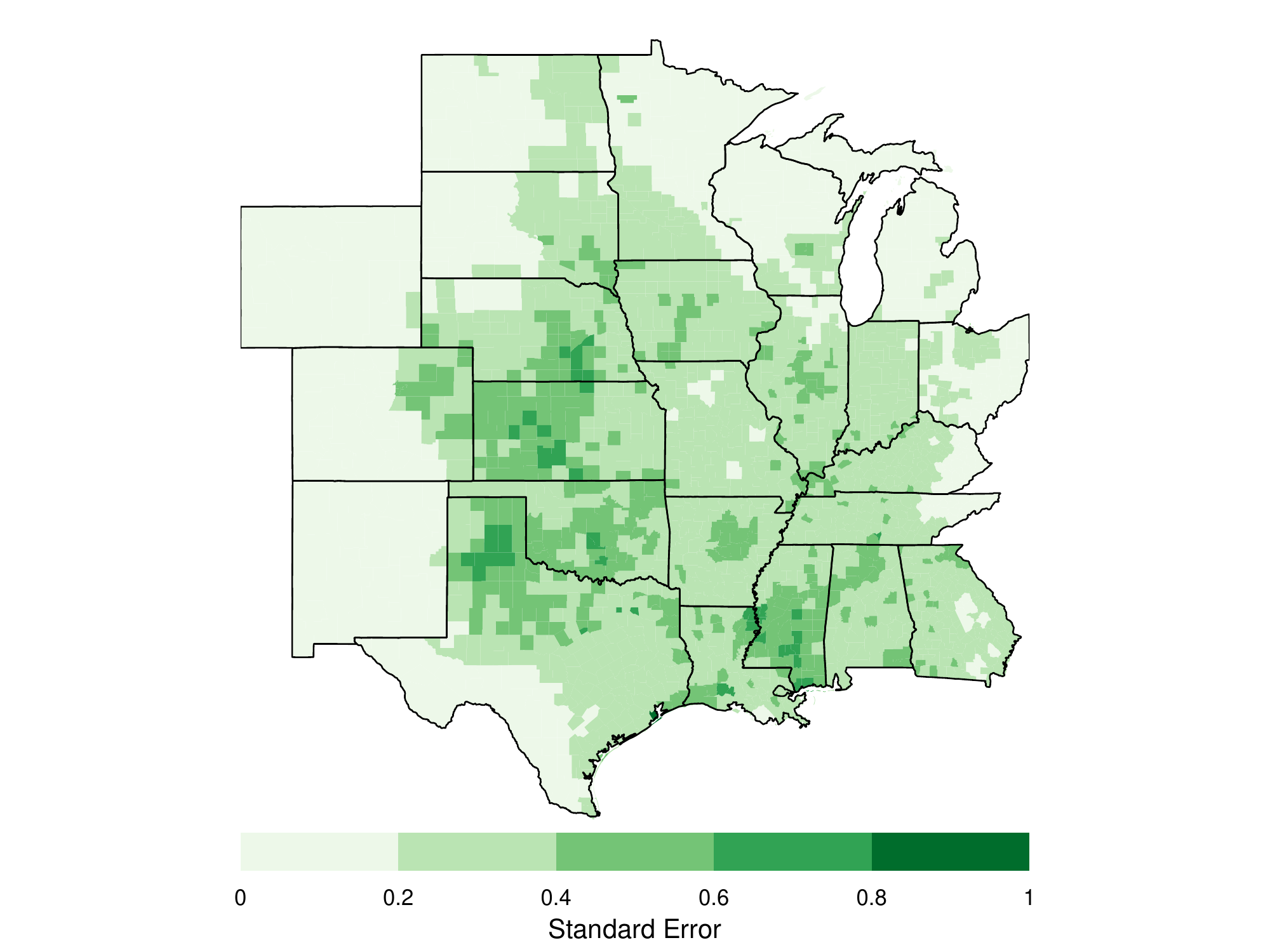}
 \caption{Variation in long-term rates. Standard deviation of tornado occurrence per 100 square kilometers.}
\label{StandardErrors}
\end{figure}

Uncertainty about the predicted risk is metered by the standard deviation of the predictive distribution (Fig.~\ref{StandardErrors}). In general, areas with the largest standard errors are found in regions with the highest rates. For most of the study domain the uncertainty amounts to less than .4 tornadoes per 100 square kilometers. Galveston County Texas has the largest uncertainty in excess of .9 tornadoes per 100 square kilometers.

\subsubsection*{Prediction Quality}

\begin{figure}[!h]
\centering
\noindent \includegraphics[width=4in]{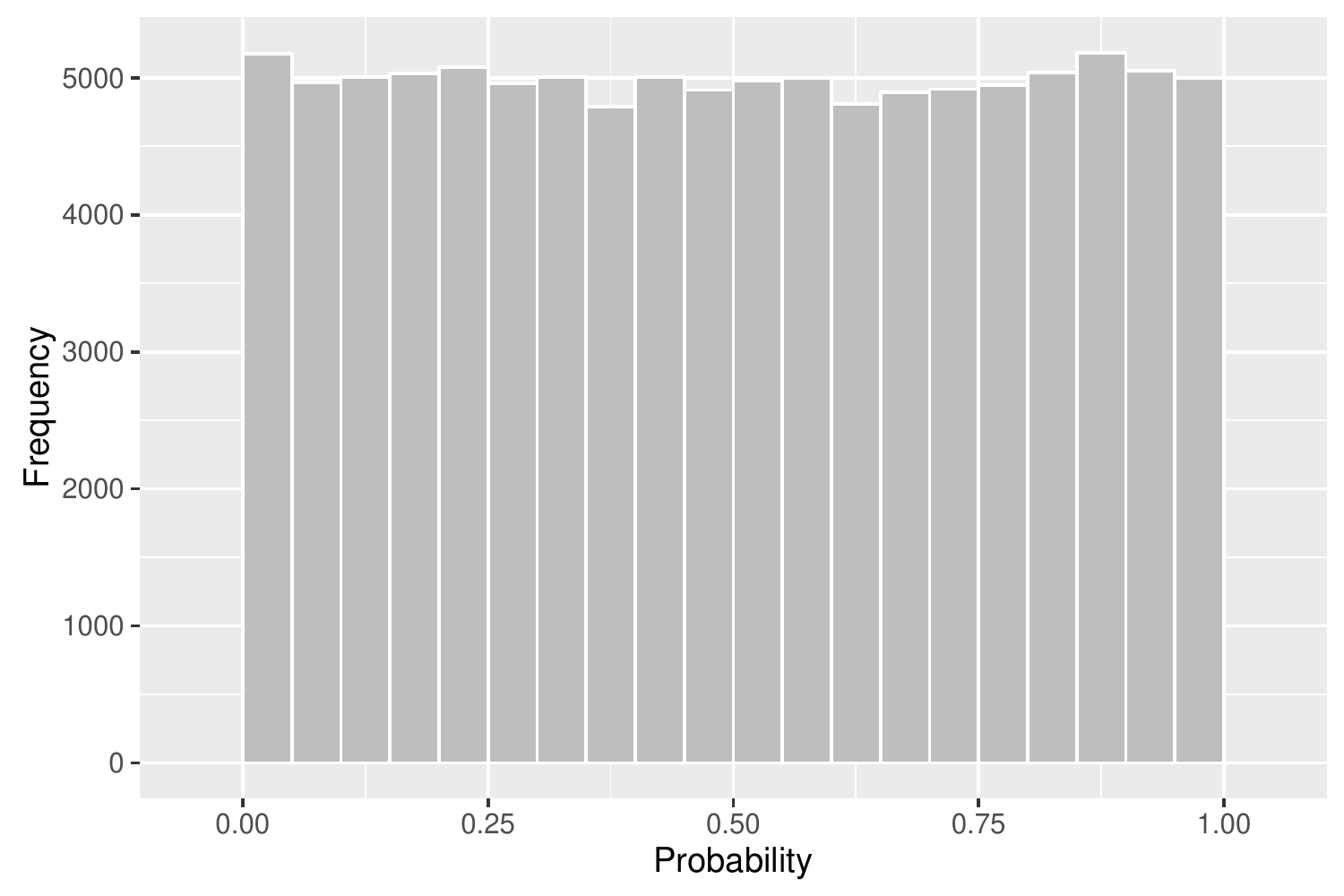}
 \caption{Distribution of the modified probability integral transform values. The values have a distribution that is looks uniform.}
\label{PIT}
\end{figure}

The predictive quality of the model is assessed by the cross-validated log score, the correlation between counts and expected rates, and the distribution of probability integral transform values. The log score is equivalent to a mean square error with smaller values indicating better prediction quality \cite{GneitingRaftery2007}. For all tornadoes the log score is .83 and the correlation between observed counts and expected rates is $+$.24. While the value is small relative to a perfect correlation the observations are counts while the predictions are rates so the upper limit is much less than unity especially given the over-dispersed counts. The distribution of the modified probability integral transform values is nearly uniform (Fig.~\ref{PIT}) indicating that the model fits the data well and that there is no model bias.

\begin{figure}[!h]
\centering
\noindent \includegraphics[width=3in]{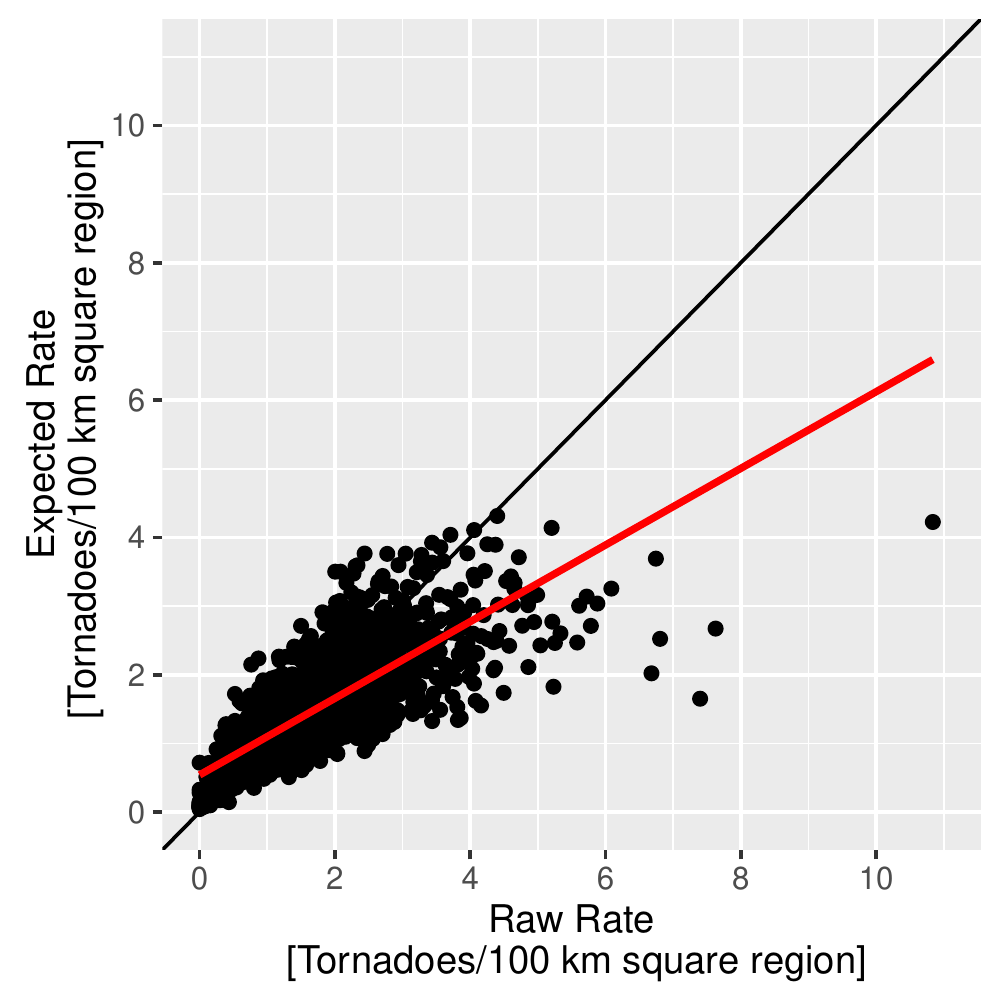}
 \caption{Raw versus expected tornado rates. The best-fit regression line is shown in red.}
\label{RawVsExpected}
\end{figure}

Comparing rates provides additional insight on the quality of the model. The raw rate is the number of tornadoes in the county divided by the number of years and scaled to have units of per 100 km squared area. The correlation between raw and expected rates is $+$.79. The relationship is shown in Fig.~\ref{RawVsExpected}. Each point on the graph indicates a county. Note that the raw rate does not control for trends or for the tendency of having fewer reports in less populated counties. Divergence between the raw and expected rates for large rates results from the spatial smoothing imposed by the spatially correlated random term. Counties that have been exceptionally unlucky in terms of tornado strikes relative to neighboring counties are smoothed toward the neighborhood average. In the absence of evidence that a county has a unique risk relative to its neighbors, this is how it should be.

\begin{figure}[!h]
\noindent \includegraphics[width=5in]{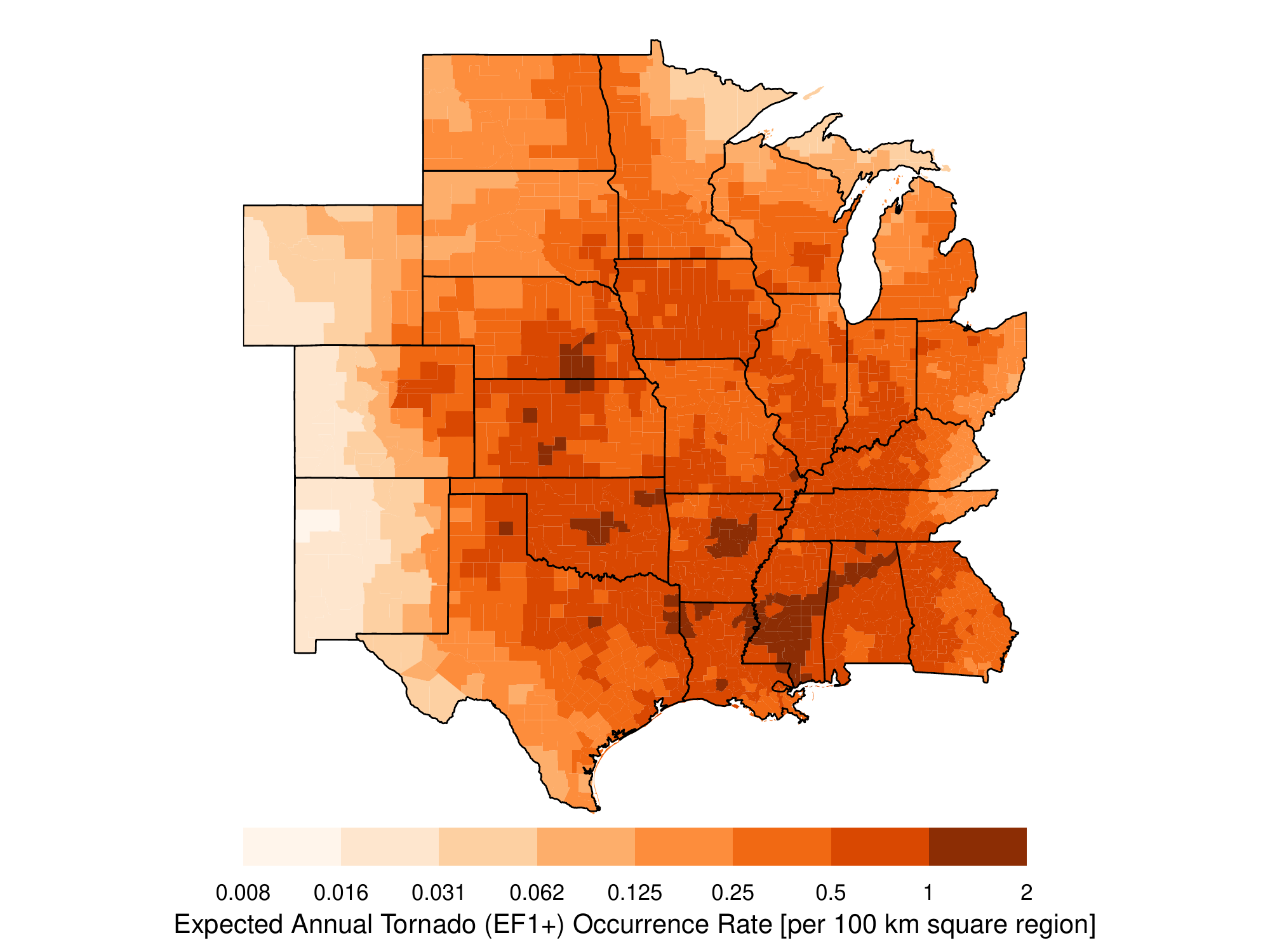}
 \caption{Expected annual tornado (EF1+) occurrence rates per 100 square kilometers.}
\label{OccurrenceRates1}
\end{figure}

A similar model is fit to a subset of data consisting only of tornadoes rated EF1 and higher. Just over two percent of the counties did not experience an EF1+ tornado over the period of record. With this subset of data the interaction term between population and year is not included because it is not statistically significant. Occurrence rates are consistently above one tornado every other year across most of the region (Fig.~\ref{OccurrenceRates1}). Highest rates occur across the mid South and central Great Plains. Relative to the model fit on all the data, the log score is improved at .58 but the correlation between observed counts and predicted rates drops to .20 as expected with many counties having no tornadoes in a given year. The distribution of the modified probability integral transform values is again nearly uniform indicating that the model fits the data well and that there is no problem with model bias. The correlation between raw and predicted rates increases to .84.

\subsection*{Short-Term View}

\subsubsection*{Diagnostic Mode}

Long-term rates provide a background climatology from which 100- or 1000-year losses can be projected given additional models for intensity and damages. The risk in a given year for a particular region however might be higher (or lower) than the long-term rate depending on climate factors so it is valuable to have a short-term view of the risk. The model described in Eqs.~(6-9) is fit to the annual tornado counts in raster cells. The linear trend term indicates an average annual increase in tornadoes over the domain of 1.2\% [(1.1, 1.3)\%, 95\% credible interval (CI)] consistent with improvements in observing practices. The posterior mean of the spatially-uniform GAK term ($\beta_3$) on the short-term view model indicates a 4.3\% [(0.6, 7.9)\%, 95\% CI] reduction in tornadoes across the domain for every one degree increase in SST in this region consistent with an earlier study over the Central Plains (\cite{ElsnerWiden2014}).

\begin{figure}[!h]
\centering \includegraphics[width=5.5in]{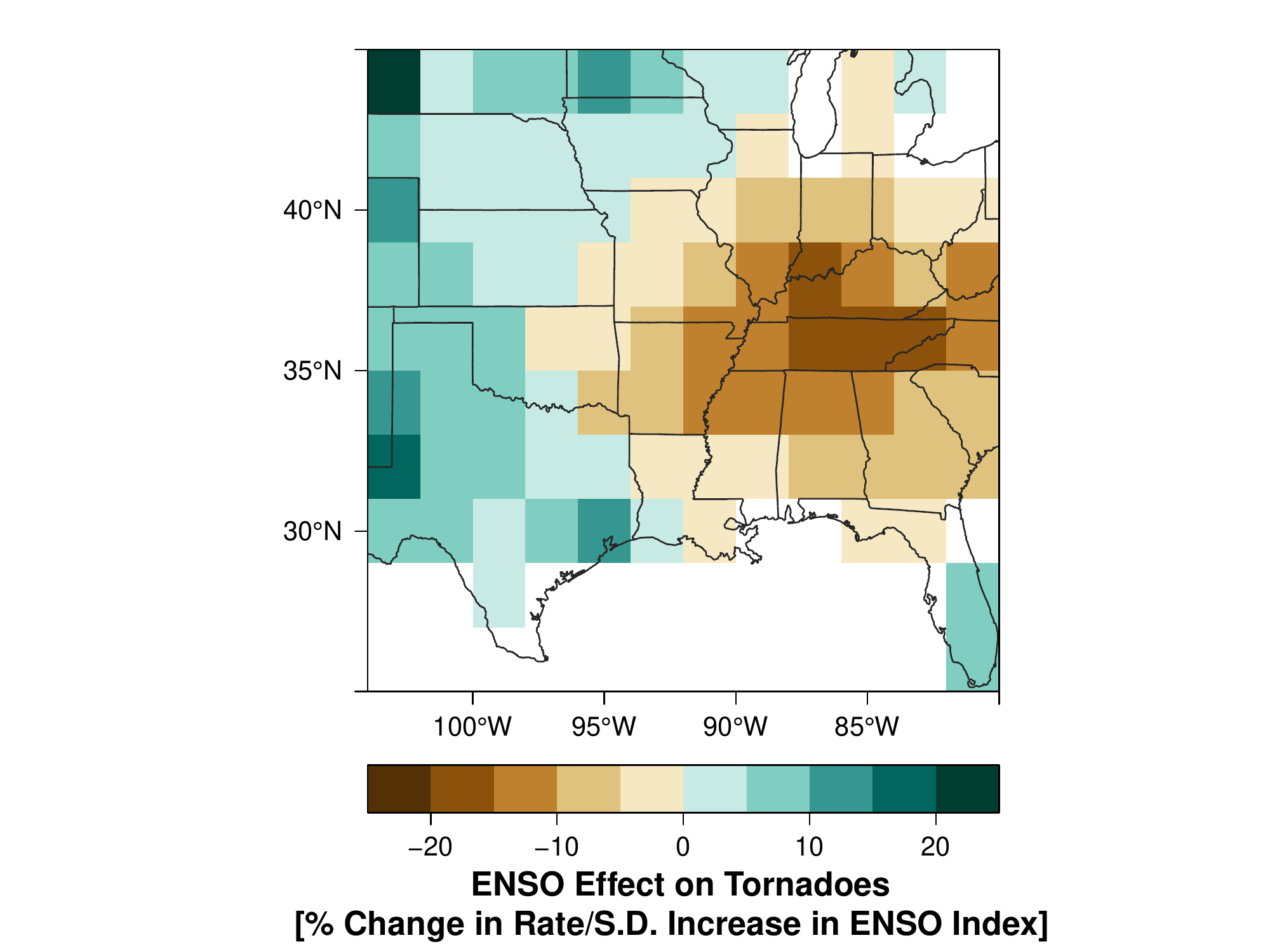}
 \caption{Short-term view. ENSO effect on tornadoes. Magnitude of the effect in units of percentage change in tornado rate per standard deviation (s.d.) increase in the springtime (Mar--May) value of the bi-variate ENSO index. Positive values indicate more tornadoes during El~Ni\~no.}
\label{ENSOeffect}
\end{figure}

The posterior mean of the spatial-varying ENSO term ($\beta_{4,s}$) answers the question: what is the geographic pattern of the ENSO effect on tornadoes? Spatially the ENSO effect implies fewer tornadoes over the Mid South and more across the High Plains during El~Ni\~no (Fig.~\ref{ENSOeffect}). The reduction exceeds 15\% over a large part of Tennessee and extends westward to northeastern Texas and southeastern Kansas and northward into eastern Wisconsin and Michigan. The enhancement in tornado activity across the Plains extends from western Texas northward to western South Dakota. The statistical significance of the effect is estimated by dividing the posterior mean by the posterior standard deviation. Areas of the Mid South centered on Tennessee show the most significance. Areas across the High Plains that tend to get more tornadoes during El~Ni\~no have lower significance levels.

\begin{figure}[!h]
\centering \includegraphics[width=5.5in]{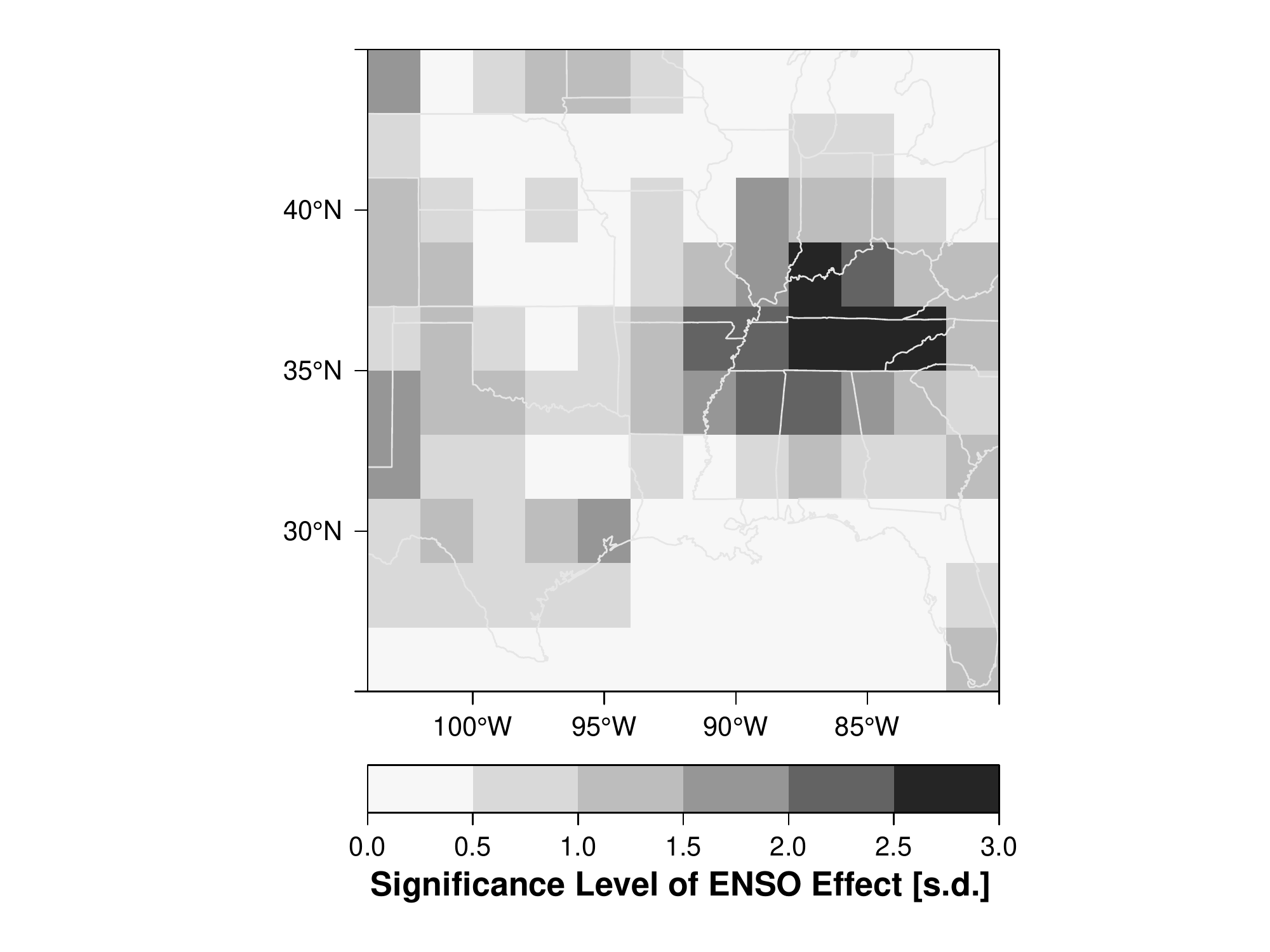}
 \caption{Significance of the ENSO effect on tornadoes in units of standard deviation.}
\label{ENSOsig}
\end{figure}

\begin{figure}[!h]
\centering \includegraphics[width=5.5in]{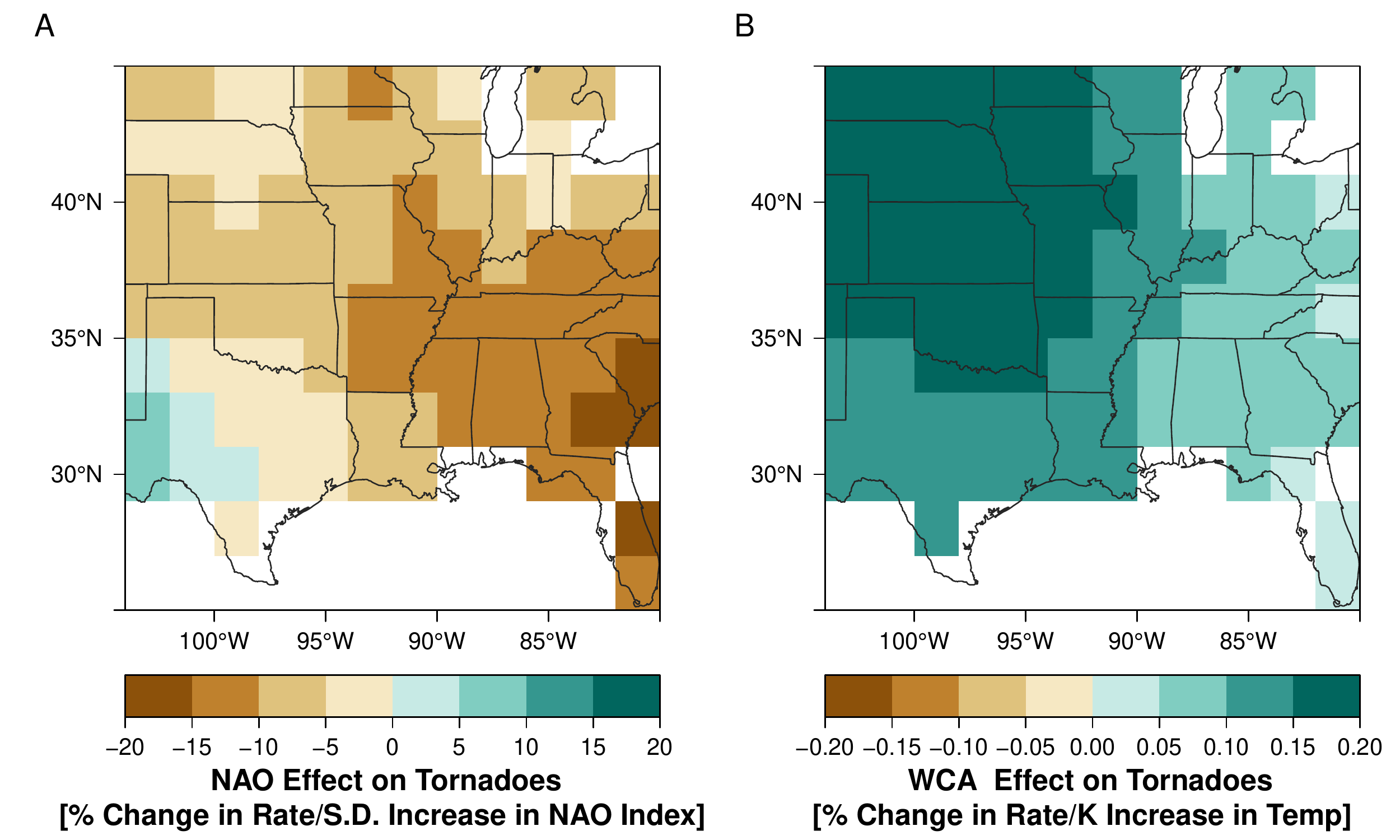}
 \caption{NAO and WCA effect on tornadoes. (A) Magnitude of the NAO effect in units of percentage change in tornado rate per s.d. increase in the springtime (Apr--Jun) value of the NAO index. Negative values indicate fewer tornadoes with a positive NAO. (B) Magnitude of the WCA effect in units of percentage change in tornado rate per $^\circ$C increase in February SST in the Western Caribbean Sea.}
\label{NAOWCAeffects}
\end{figure}

The NAO effect is similar to the ENSO effect in magnitude although not as symmetric (many more cells have negative values) with most of the Southeast extending into the Central Plains indicating a decrease in tornado occurrence with a positive NAO index (Fig.~\ref{NAOWCAeffects}). Only portions of West Texas indicate an increase in tornado frequency with a positive NAO. A positive NAO is marked by lower heights (or pressures) over Iceland and higher heights over the subtropics. Not surprisingly the magnitude of the effect is most pronounced over Georgia and South Carolina closest to the mean position of the subtropical high pressure area. The WCA effect indicates a general increase in the probability of tornadoes especially over the normally drier regions of the central and northern Great Plains but the magnitude of the effect is considerably smaller than the ENSO and NAO effects.

\subsubsection*{Predictive Mode}

\begin{figure}[!h]
\centering \includegraphics[width=5.5in]{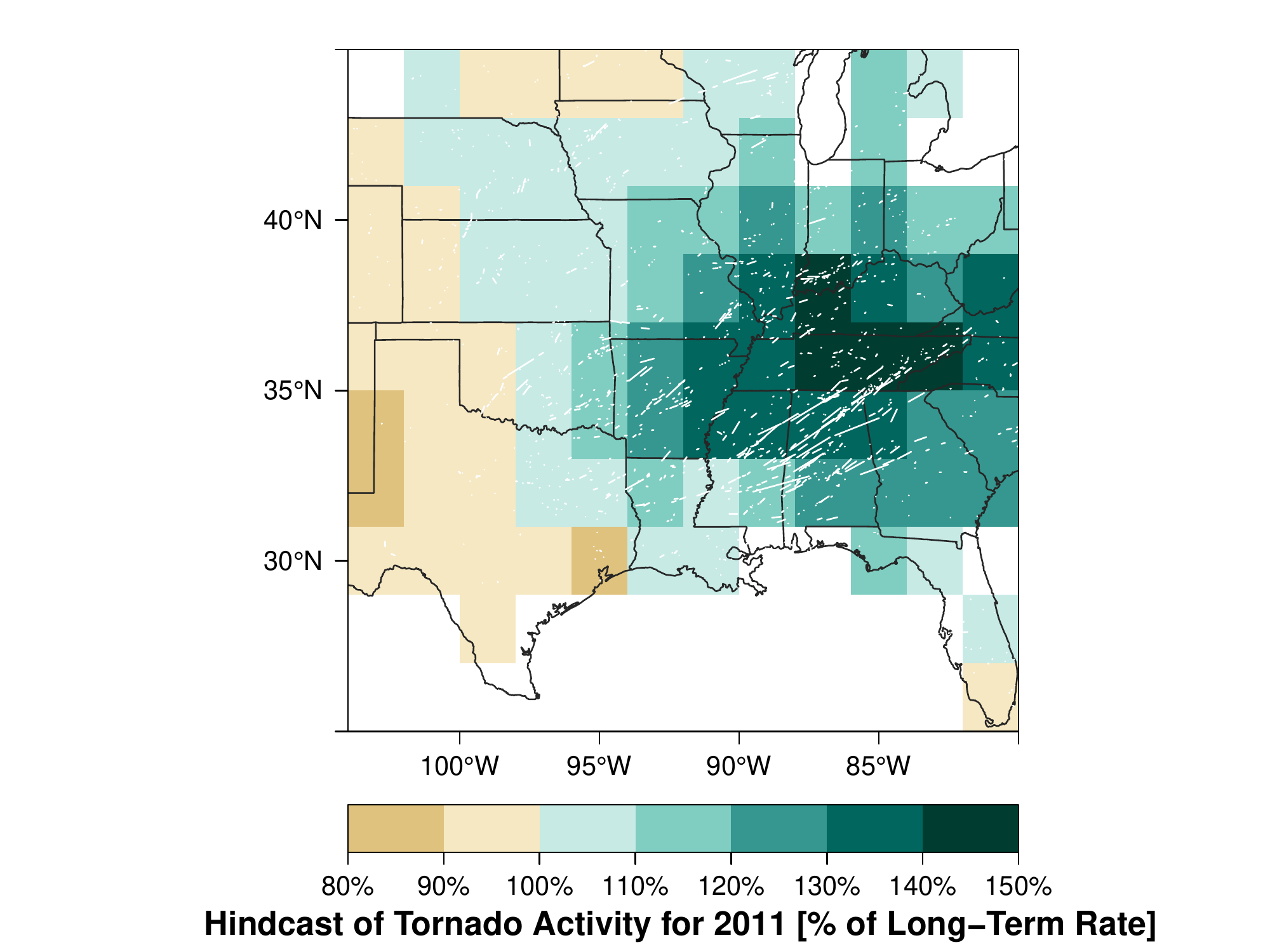}
 \caption{Hindcast of the 2011 tornado year. The straight-line tracks of all tornadoes during that year are shown in white.}
\label{2011prediction}
\end{figure}

A short-term view is a hedge against the long-term rates when climate signals are strong and effects are aligned. The results from the short-term model run in a diagnostic mode indicate how to adjust rates on average during El~Ni\~no holding the other climate variables constant. But to quantify the {\it combined} effect of the predictors the short-term view model is run in predictive mode. To predict the rate adjustment for a particular year, the cell counts for that year are left out of the model fit. Because the approach is Bayesian where all parameters and data are treated as random variables, the fitting procedure estimates the cell rates for the year removed. As an example, early in 2011 a La~Ni\~na event was occurring and the springtime NAO index was negative. Together with lower than normal Gulf of Alaska SST the stage was set for an increased threat of tornadoes across the Southeast. Removing the counts for 2011 the model predicts (hindcasts) increased tornado activity across a large part of the region east of 97$^\circ$ W longitude with much of Tennessee expecting rates to be near 150\% of the long-term rate (Fig.~\ref{2011prediction}). The straight-line tracks of all tornadoes during 2011 are shown as white lines clearly indicating the predicted preference for the activity across the Southeast. A portion of the 2011 tornadoes occurred before the months used to create the predictor indexes so the map does not represent a true forecast.

\section*{Discussion and Conclusion}

This paper demonstrates that regionally specific seasonal tornado forecasts are possible with spatial statistical models. The models are built from data in the historical record. Modern fitting techniques are able to handle clustered and pathological records. The models produce long- and short-term views of regional tornado risk. The most important predictor examined is ENSO. During spring in a La Ni\~na phase of ENSO a strengthened Inter-American Seas (IAS) low-level jet enhances the spread of moisture across the Southeast United States. Greater instability associated with more moisture is coupled with increased shear from a strengthened upper-level jet setting the stage for more tornadoes. Whether or not tornado activity actually increases above the long-term rate for a particular La Ni\~na depends on additional unknown and unpredictable factors but given these conditions over many cases the data show statistical evidence for an elevated risk. 

A positive phase of the NAO decreases the chance of tornadoes across a large part of the study area but especially over the Southeast. A positive NAO is associated with a strong subtropical high pressure zone over the North Atlantic that inhibits deep convection across this part of the country. The Gulf of Alaska SST term, which indicates fewer tornadoes with a warmer ocean, does not have a significant spatial component while the Caribbean Sea SST term, which indicates more tornadoes with a warmer ocean, has only weak spatial variability.

The long-term rates can be used by property insurance companies to set policy rates and by emergency managers to allocate resources. The short-term rate adjustments can be used by reinsurance companies and hedge funds looking for ways to adjust a risk portfolio. Future work will focus on fitting the models to the historical database of Grazulis \cite{Grazulis1990}. The additional data will enhance the precision on the long-term rate estimates and will better define the influence climate predictors have on the short-term rates. Also, since the predictors in the short-term view model were based on previous studies, future work can focus on a more comprehensive examination of other climate variables. The code for the long-term view model is available at \url{rpubs.com/jelsner/tornadoRisk_longTermView} and can be modified to estimate risk in other tornado-prone states and regions.

\section*{Acknowledgments} This work was supported by the Risk Prediction Initiative of the Bermuda Institute for Ocean Studies (BIOS) with special thanks to Mark Guishard and John Wardman. The authors declare no competing interests.

\nolinenumbers

\bibliography{References.bib}
\bibliographystyle{plos2015.bst}

%
\end{document}